\documentstyle[multicol,prl,aps,psfig]{revtex}

\begin{document}

\draft
\title{Mechanisms for electron transport in atomic-scale \\ one-dimensional
wires: soliton and polaron effects}
\author{
H. Ness \footnote{\ Present address: 
CEA-Saclay, DSM/DRECAM/SPCSI, B\^at. 462, F-91191 Gif-sur-Yvette, France.
E-mail: ness@cea.fr}
and 
A.J. Fisher \footnote{\ E-mail: Andrew.Fisher@ucl.ac.uk}}
\address{Department of Physics and Astronomy, University College London,\\
Gower Street, London WC1E 6BT, U.K.}
\maketitle

\begin{abstract}
We study one-electron tunneling through atomic-scale one-dimensional
wires in the presence of coherent electron-phonon ($e$-ph) coupling.
We use a full quantum model for the $e$-ph interaction within the
wire with open boundary conditions.
We illustrate the mechanisms of transport in the context of molecular
wires subject to boundary conditions imposing the presence of a
soliton defect in the molecule.
Competition between polarons and solitons in the coherent transport
is examined.
The transport mechanisms proposed are generally applicable to other
one-dimensional nanoscale systems with strong $e$-ph coupling.
\end{abstract}

\pacs{PACS numbers: 73.63.Nm, 73.40.Gk, 73.61.Ph}

\begin{multicols}2

\section{Introduction}

How will an electronic system conduct in the limit where it becomes
entirely one-dimensional?  This became an issue in the late 1980s,
with the fabrication of semiconductor structures that possessed only a
single conducting ``channel'', and which exhibited the phenomenon of
quantized conductance \cite{quantcond88}.  With nanoscale fabrication
techniques, the class of effectively one-dimensional systems has
expanded to include: atomic-scale structures on surfaces produced by
scanning probe lithography \cite{dbwire}, self-assembly
\cite{metalonsi}, or directed assembly \cite{BW+MA}; break junctions
\cite{bjncs,ohnishi98}; carbon nanotubes \cite{lemay01}; and
conjugated organic molecules \cite{reed97}.  The measurement of the
transport properties of individual examples of such systems is now
possible; for example, conduction in individual molecules has recently
been measured by scanning probes \cite{langlais99,donhauser01} or bulk
electrodes \cite{reed97,kergueris99}; transport in nanotubes has been
measured using lithographed electrodes \cite{tans98}.
Similar measurements may also soon become possible for atomic-scale 
wires on surfaces.

Many of these recent examples approach the ideal of a system whose
atomic---as well as electronic---structure is truly one-dimensional.
As such they may be expected to exhibit enhanced effects of
correlations during electron transport.  Electron-electron
correlations have consequences including the possible formation of
Luttinger liquids \cite{dash01}; on the other hand, the effects
produced by electron-atom correlations can include the Peierls
transition \cite{peierls}, which opens a gap at the Fermi energy and
renders one-dimensional metallic systems semiconducting.  If these
quasi-one-dimensional systems are to have applications in
nanotechnology, it is vital to understand whether currents can pass
through them, despite the possibility of a Peierls transition.  In
this paper we address this question, in the context of molecular
wires; specifically, we consider finite-length poly-acetylene chains
described by a fully quantum version of the Su--Schrieffer--Heeger
(SSH) model \cite{heeger88}.

We have chosen to study molecular conductors here, rather than any of
the other types, because electron-lattice coupling in bulk samples is
relatively well studied \cite{heeger88}.  The Peierls transition
manifests itself in an alternation of single and double bonds (i.e.,
{\it dimerisation}) along the molecule's length, and bulk electron
transport is dominated by mobile intrinsic defects formed when
electrons or holes are injected into this dimerised structure.  For
our purposes the most important defects are polarons (which can be
thought of as a local reduction in the dimerisation around an injected
charge) and charged solitons (topological defects in the bond length
alternation).  The soliton has an associated electronic state at
mid-gap and the dimerisation changes sign, passing through zero in the
neighbourhood of the defect \cite{heeger88}.

In order to understand molecular charge transport on nanometre
lengthscales, we must develop a theory of coherent transport that
accounts for polaron and soliton formation.  We took the first steps
recently when we showed that the tunneling of carriers through
molecular wires at low temperature is dramatically enhanced by the
formation of virtual polarons \cite{ness}.  
Therefore, previous treatments of tunneling in molecules based
on elastic scattering significantly underestimate the conductance
\cite{elastrans}.
However, the boundary conditions used in our
previous work prevented solitons from forming in the molecule, as the
dimerisation was constrained to be equal at both ends.  In the present
paper we study chains with an odd number of monomers; the ground state
of such a system possesses a single soliton in the centre
\cite{heeger88,lubook}.
We are thus able to study for the first time the role of the moving 
soliton, with its associated mid-gap state, and the competition between 
polarons and solitons in coherent transport.

\section{Physical model}

The model for the molecular wires includes delocalized $\pi$-electrons
interacting with quantum phonons.
The electronic Hamiltonian is expressed in the basis of the  
one-electron eigenstates (labelled by $n,m$) of the 
reference
system. The {\it e}-ph coupling is linear in the phonon displacements
and induces transitions between electronic states.
The 
molecule Hamiltonian is
\begin{equation}
H=\sum_n\epsilon_n c^\dag_n c_n
+\sum_\lambda\hbar\omega_\lambda a^\dag_\lambda a_\lambda
+\sum_{\lambda,n,m}\gamma_{\lambda nm}(a^\dag_\lambda+a_\lambda)
c^\dag_n c_m\ ,
\label{Hamilt}
\end{equation}
where $c^\dag_n$ ($c_n$) creates (annihilates) an electron in the
$n$-th electronic state with energy $\epsilon_n$ and $a^\dag_\lambda$ 
($a_\lambda$) 
creates (annihilates) a quantum of energy $\hbar\omega_\lambda$ in the 
eigenmode of vibration $\lambda$ of the isolated molecule.
The values of $\epsilon_n$, $\omega_\lambda$ and 
$\gamma_{\lambda nm}$ are 
calculated from the ground state 
of the neutral molecule (described by the SSH model  \cite{heeger88,ness})
containing an odd number $N$ of monomers.
The vibrational eigenmodes $V_\lambda$ and frequencies of the molecule 
are calculated within the harmonic approximation \cite{chao85}.

To calculate the transport properties through the wire, the 
left and right ends of the molecule (atomic sites $i=1$ and $i=N$) 
are connected to metallic 
leads via hopping integrals $v_{L,R}$ respectively.
The leads are modeled as one-dimensional semi-infinite tight-binding
chains with on-site energy $\epsilon_{L,R}$ and inter-site hopping
matrix elements $\beta_{L,R}$ (with no $e$-ph coupling).
The scattering states $\vert\Psi\rangle$ for a single incoming charge 
carrier are expanded inside the molecule onto the eigenstates
$\vert n,\{n_\lambda\}\rangle=c^\dag_n\prod_\lambda 
(a^\dag_\lambda)^{n_\lambda}/\sqrt{n_\lambda!}\ \vert 0\rangle$,
of the non-interacting $e$-ph system
\footnote{\ A similar expansion is used in the leads, $c^\dag_n$ is then 
replaced by the charge creation operator $d^\dag_i$ on site $i$.}.
$n_\lambda$ is the occupation number of the phonon mode $\lambda$ and 
$\vert 0\rangle$ is the {\it vacuum} state (representing the neutral ground
state 
of the whole system, with a definite number of electrons in each part).  
The single added carrier can 
be anywhere in the system 
(on the left, the right, or in the molecule) and interacts with
phonons only when inside the molecule.  We only consider
current-carrying states in which a single electron is added to the
ground state; we expect this assumption to hold since the interval
between electron transmission is much greater than the transit time
\cite{ness}.

We now briefly outline the calculation technique.
The transport problem is solved by mapping the many 
body (one-electron/many-bosons) problem onto a single-electron problem 
with many scattering channels \cite{trugman,ness}.
Each channel represents the different scattering processes by which 
the electron might exchange energy with the phonons.
For an initial phonon distribution $b\equiv\{m_\lambda\}$ and 
an incoming electron from the left lead, the outgoing channels in the left 
and right leads are associated with energy-dependent reflection 
$r_{ab}(\epsilon)$ and transmission $t_{ab}(\epsilon)$ coefficients 
respectively ($a\equiv\{n_\lambda\}$ is the phonon
distribution after scattering in each outgoing channel).
In the leads, the scattering states take the asympotic form of propagating
Bloch waves 
with amplitudes $r_{ab}$ (reflection) and $t_{ab}$ (transmission). 
The wave vectors are
given by the dispersion
relations in each channel.
For example, the energy $\epsilon_{in}$ of the incoming electron from
the left lead is related to the wave vector $k_b^L$ by the 
tight-binding-like dispersion relation 
$\epsilon_{in}=\epsilon_L+2\beta_L\cos k_b^L$.
For the final energy $\epsilon_{fin}$ of the electron transmitted 
to the right, one has:
$\epsilon_{fin}=\epsilon_R+2\beta_R\cos k_a^R$.
One can then project out the leads from the problem and work 
in the molecular wire subspace to solve for the value of the scattering
state $\vert\Psi(E)\rangle$.
This state is obtained from propagating the source term
$\vert s(E)\rangle$ (incoming electron from the left)
via the effective Green's function $G(E)$ defined in the molecular wire 
subspace: 
$\vert\Psi(E)\rangle=G(E)\vert s(E)\rangle$, where $G$ is given
by
$G(E)=[E-H-\Sigma_L(E)-\Sigma_R(E)]^{-1}$,
$H$ is the molecular wire Hamiltonian defined in Eq. (\ref{Hamilt}) and
$\Sigma_{L,R}(E)$ are complex potentials arising from embedding the
molecular wire spectrum into the continuum of states associated
with the leads. 
Overall energy is conserved, so $\epsilon_{in}$ 
and $\epsilon_{fin}$ are related by: 
$E=\epsilon_{in}+\sum_\lambda m_\lambda\hbar\omega_\lambda
=\epsilon_{fin}+\sum_\lambda n_\lambda\hbar\omega_\lambda$
\footnote{\ In this paper, we consider the limit of low temperatures:
the initial phonon distribution is $b\equiv\{m_\lambda\}=\{0\}$ where 
all optic phonon modes are in the ground state
(in this limit, the injection energy equals the total 
energy 
$\epsilon_{in}=E=\epsilon_{fin}+\sum_\lambda n_\lambda\hbar\omega_\lambda$).
It is a very good approximation even at room temperature for all optic
modes except the soliton translation.
For the soliton translation the condition $kT\ll\hbar\omega$ restricts us
to $T\ll 230K$.}.

The linear system $\vert\Psi\rangle=G\vert s\rangle$ is solved for
a finite size basis set by truncating the phonon subspace ({\it i.e}
considering the lowest occupation numbers up to $n_{\rm occ}^{\rm max}$ 
in each mode).
Furthermore the electron is coupled inside the molecule to a finite
number $N_{\rm ph}$ of the most relevant phonon modes (see next
section).   
A detailed analysis of the validity of such approximations can be
found in Ref. \cite{ness}.
From the solution $\vert\Psi\rangle=G\vert s\rangle$,
one can obtain the 
reflection $r_{ab}$ and transmission $t_{ab}$ coefficients 
for all the 
channels and 
hence the currents flowing through the wire.
One can also calculate the expection values of any 
{\it correlation functions} between the electron 
and phonon degrees of freedom.

We define the transmission
probability $T_{ab}=\vert t_{ab}(\epsilon_{in})\vert^2
\beta_R\sin k_a^R(\epsilon_{fin})/(\beta_L\sin k_b^L(\epsilon_{in}))$
\footnote{\ $T_{ab}$ contains as usual the ratio of the velocity in
the outgoing channel to that in the incoming
channel.}.
In our model, $\vert t_{ab}\vert^2$ is proportional to 
$\vert\langle i=N, a\vert\Psi\rangle\vert^2=
\vert\langle N, a\vert G\vert s\rangle\vert^2$
and the source term is proportional to the velocity of the
incoming electron \cite{ness}.
Then, working in the real-space representation,
the inelastic transmission probability can 
be rewritten in the following usual form \cite{sols92} :
\begin{eqnarray}
T_{ab}(\epsilon_{fin},\epsilon_{in})
=4\ \frac{v_L^2}{\beta_L}\sin k_b^L(\epsilon_{in})\  
  \frac{v_R^2}{\beta_R}\sin k_a^R(\epsilon_{fin})\ \nonumber \\
  \times \vert\langle i=N\vert G_{ab}(E)\vert i=1\rangle\vert^2 \ ,
\label{Tinel}
\end{eqnarray}
where 
$\langle N\vert G_{ab}(E)\vert 1\rangle$ is the matrix element of
the Green's function $G$ taken between the 
left side $i=1$ and the right side $i=N$ of the molecule and 
the phonon configurations before ($b$) and after ($a$) scattering
\footnote{\ The factors 
${v_{L,R}^2}/{\beta_{L,R}}\sin k_{b,a}^{L,R}$ in Eq.(\ref{Tinel}) are 
related to the imaginary parts of the potentials $\Sigma_{L,R}$.}.

\section{Results}

\begin{figure}[p]
\begin{center}
\psfig{figure=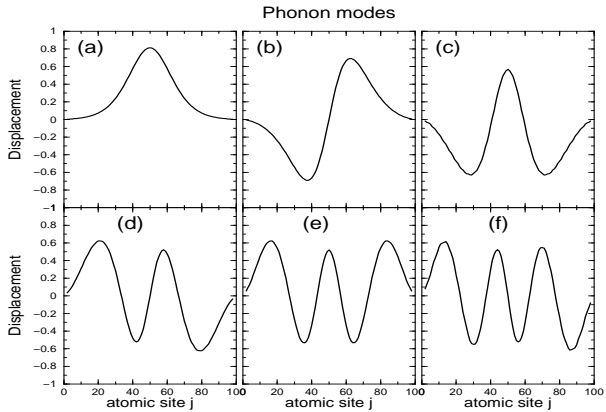,width=8cm,height=5.5cm}
\end{center}
\narrowtext
  \caption{\label{fig1}
Optic component of the phonon modes 
for a chain length $N=99$.
The optic component is defined as: 
$(-1)^j(V_\lambda(j+1)-2V_\lambda(j)+V_\lambda(j-1))$.
(a) is the mode associated with the soliton translation
($\hbar\omega_{\rm a}$=0.020 eV),
(b) with the deformation of the soliton width 
($\hbar\omega_{\rm b}$=0.114 eV), 
(c) is the so-called third mode ($\hbar\omega_{\rm c}$=0.134 eV). 
(d,e,f) are other long-wavelength optic modes 
($\hbar\omega_{\rm d,e,f}$=0.144, 0.152, 0.158 eV).}
\end{figure}

\begin{figure}[p]
\begin{center}
\psfig{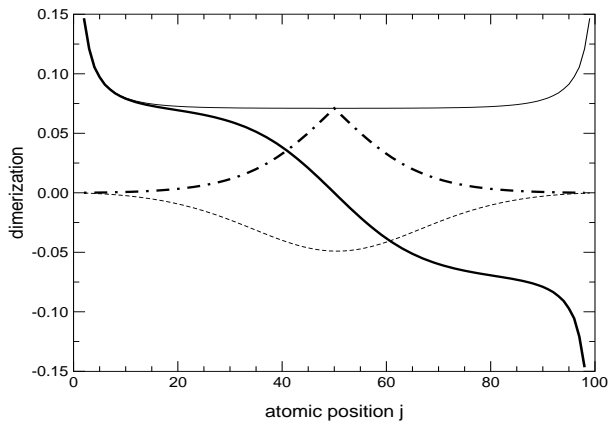}
\end{center}
\narrowtext
  \caption{\label{fig2}
Dimerisation $d_j$ (in \AA) for isolated neutral molecules.
{\it Thin solid line}: $N=100$, for even $N$, $d_j$ is constant 
($d_c$) in the middle of the chain.
{\it Solid line}: $N=99$, for odd $N$, a soliton exists in 
the middle of the chain separating 
two domains of dimerisation with opposite site.
{\it Dot-dashed line}: the absolute value $\vert d_j\vert$ 
from which $d_c$ and the end effects are substracted ($N=99$).
{\it Thin dashed line}: deformation induced by adding an extra 
electron in the chain $N=100$; 
the local reduction of the dimerisation represents a polaron.}
\end{figure}

First we describe the vibrational modes needed to obtain the 
relevant distortions of the molecule.
We have shown that polaron formation inside the molecule
after charge injection is due to the coupling 
to the long-wavelength optic modes \cite{ness}.
For odd-number chains where a soliton defect is present, we have to include
in addition the modes 
responsible for the motion and the deformation of the soliton.
We therefore include the soliton translation
(the `Goldstone mode'), the `amplitude mode'
related to the deformation of the soliton width, and higher-order 
deformation modes  (for example the so-called
third mode) \cite{heeger88,sol_mode}.
Figure \ref{fig1} shows the phonons considered for
a molecule of length $N=99$. 
As an illustrative example, we also show in Fig. \ref{fig2}
the dimerisation $d_j$ for neutral isolated even- and odd-length 
molecules.
The dimerisation is obtained, in terms of atomic displacements
$u_j$, from the staggered difference between
adjacent bond lengths: $d_j=(-1)^j(u_{j+1}-2u_j+u_{j-1})$.
For even $N$, the neutral molecule is perfectly dimerized: the 
dimerisation $d_c$ is constant in the middle of the 
chain despite the end effects.
For odd $N$, a soliton appears in the
middle of the chain, acting as a domain wall separating the
two domains of opposite sign of dimerisation.

Now we turn on the lattice deformations induced by electron
propagation in the (odd $N$) molecule connected to the 
electrodes
\footnote{\ Calculations were performed for an electron
coupled to the $N_{\rm ph}=5$ lowest frequencies modes shown
in Fig. \ref{fig1} and $n_{\rm occ}^{\rm max}=4$.
Calculations were also done for other sets of parameters.
The same qualitative physics remain when the results are
converged versus the basis set size.}.
The lattice deformations induced by the tunneling electron
are given by the correlation function
$\delta_\lambda^{[i]}$ between the electron density $P_i=c^\dag_i c_i$
on site $i$ and the displacement 
$\Delta_\lambda=(a_\lambda+a^\dag_\lambda)\sqrt{\hbar/2M\omega_\lambda}$
of the mode $\lambda$ as
$\delta_\lambda^{[i]}=\langle P^\dag_i \Delta_\lambda P_i\rangle/
\langle P_i\rangle$.
We also define the distortion on site $j$ 
due to the displaced modes when the electron is on site $i$ as 
$u_j^{[i]}=\sum_\lambda\delta_\lambda^{[i]} V_\lambda(j)$.
The dimerisation pattern $d_j^{[i]}$ is then calculated
from $u_j^{[i]}$.

\begin{figure}[p]
\begin{center}
 \leavevmode
 \psfig{figure=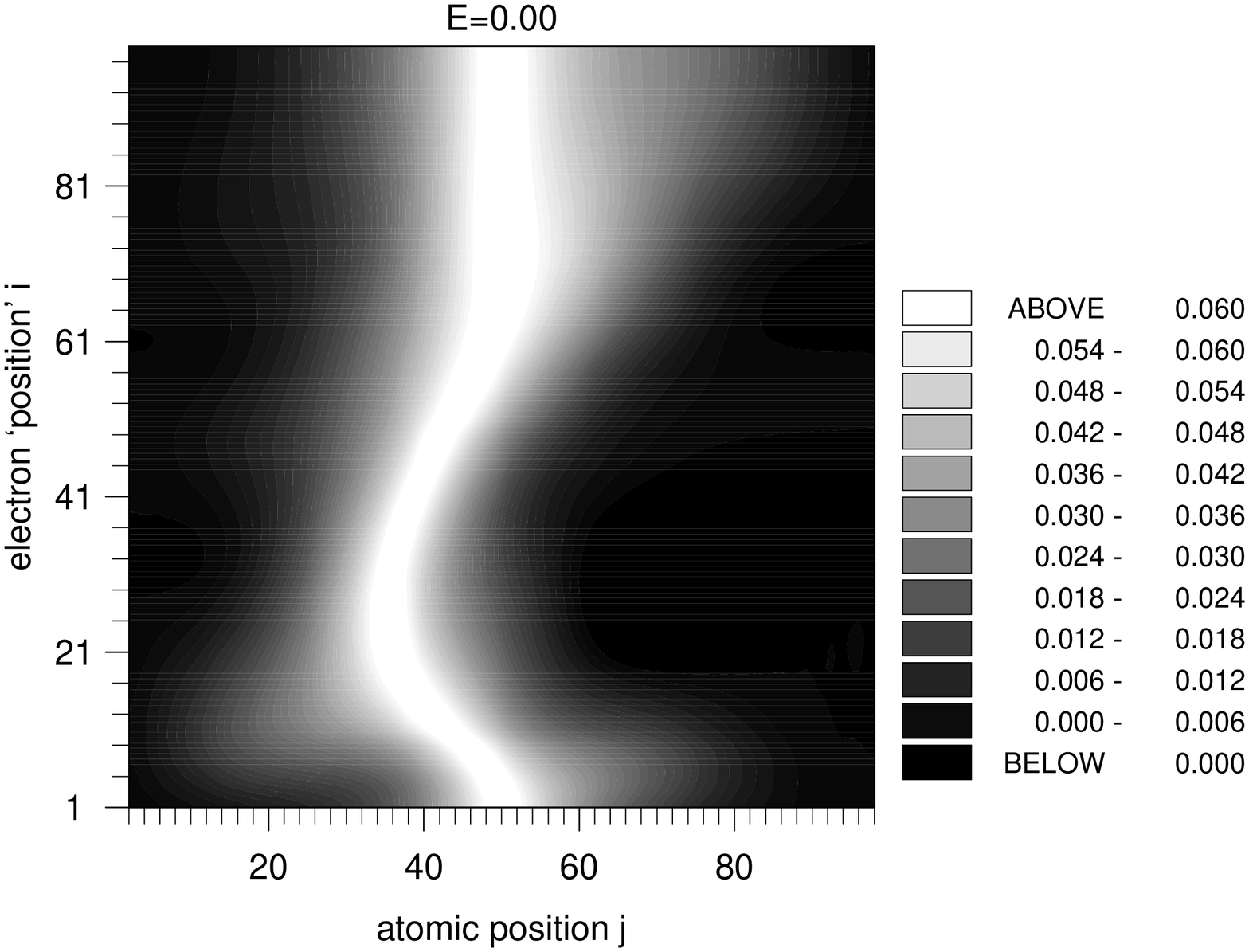,width=7.9cm,height=5.5cm}\\
  \hbox to \textwidth{\vspace*{1mm}\hfil Fig.\ \ref{fig3}(a)\hfil\hfil\hfil}
 \hspace*{1.0mm}
 \psfig{figure=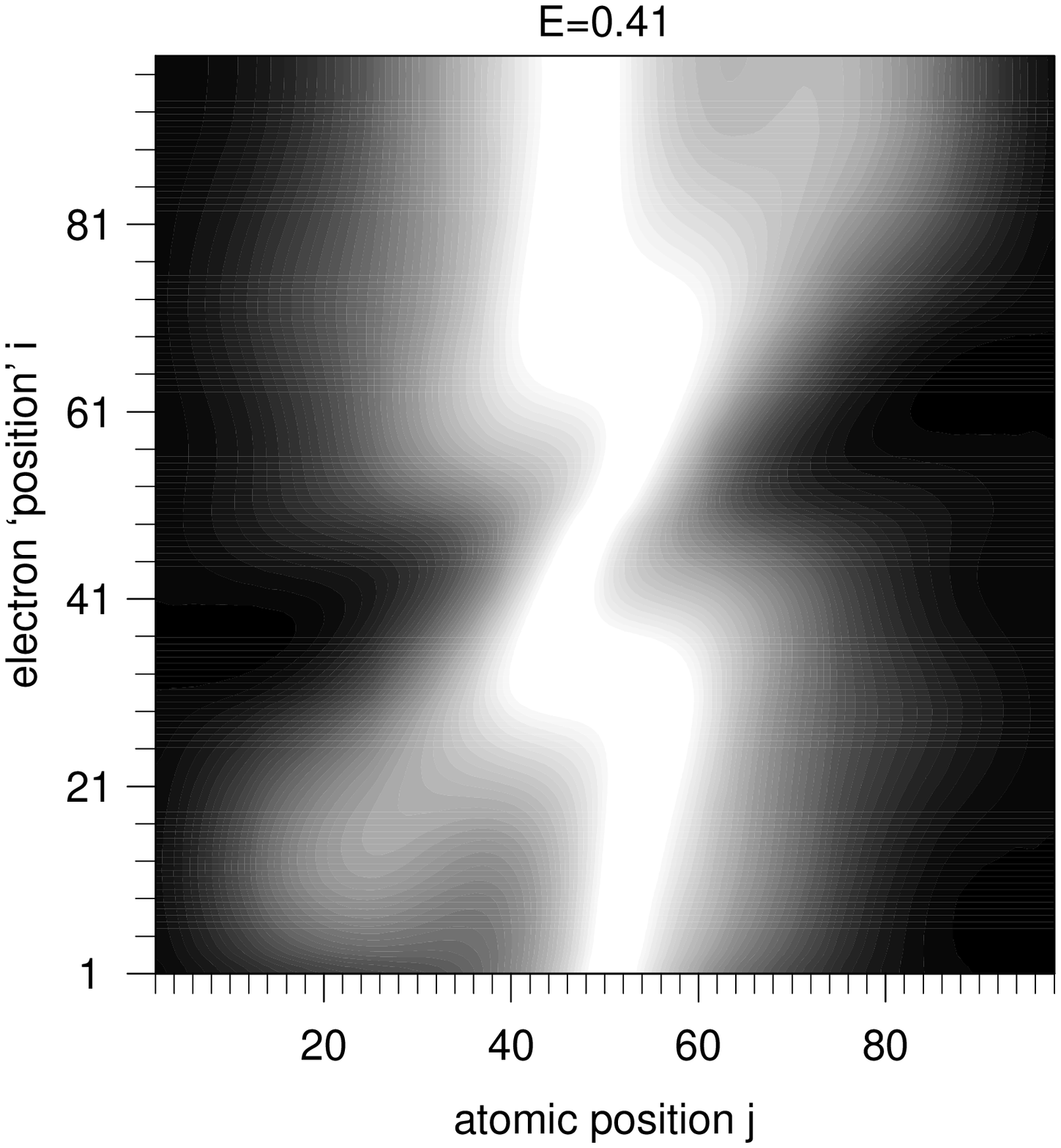,width=5.68cm,height=5.5cm}\\
  \hbox to \textwidth{\vspace*{1mm}\hfil Fig.\ \ref{fig3}(b)\hfil\hfil\hfil}
\end{center}
\narrowtext
  \caption{\label{fig3}
Two-dimensional (atomic $j$/electron $i$ positions) map of the 
dimerisation $d_j^{[i]}$ (in \AA)
obtained from the atomic distortions $u_j^{[i]}$ for the 
$N=99$ wire length.
Here we plot the absolute value 
$\vert d_j^{[i]}\vert$ substracting $d_c$ and the end effects.
The incoming electron propagates from the left to the right.
(a) Injection at mid-gap ($E=0.0$):
the bright feature represents the soliton whose position and
width vary for the different electron positions $i$ taken to
calculate $d_j^{[i]}$.
(b) Injection at $E=0.41$ eV close the polaron resonance peak in the 
transmission (Fig. \ref{fig5}):
a virtual polaron is formed (bright feature across the first
diagonal).
The maximum change in $\vert d_j^{[i]}\vert$ is $\approx 0.03-0.04$ \AA\ 
around the polaron and $\approx 0.07$ \AA\ around the soliton.}
\end{figure}

We plot on Fig. \ref{fig3} the absolute value of the dimerisation
$d_j^{[i]}$ from which the constant dimerisation $d_c$ and the end
effects have been subtracted (the corresponding pattern for the
isolated molecule is shown on Fig. \ref{fig2}).  Such a choice 
permits us to represent the defects with more contrast.  For
an injection energy $E=0$ at mid-gap, the corresponding dimerisation
is shown on Fig. \ref{fig3}(a).  The bright feature around the middle
of the chain represents the soliton.  In the absence of $e$-ph
coupling, this would correspond to charge injection in resonance with
the one-electron soliton level (see the transmission curve below).
The soliton would then be immobile, its position fixed in the
middle of the chain and its width unchanged (i.e. straight vertical
feature around $j=50$).  However due to the $e$-ph coupling, the
soliton defect becomes mobile and its width varies
while the electron progates.  Similar behaviour
is observed for other injection energies around mid-gap.
As $E$ approaches the valence band edge 
the mechanisms become different, as shown on Fig. \ref{fig3}(b).
In parallel with the soliton delocalization and deformation,
one observes the formation of a polaron.
This is characterized by the extra bright feature around the first 
diagonal in Fig. \ref{fig3}(b).
In principle, the (virtual) polaron corresponds to a local reduction
of the dimerisation around the (tunneling) electron \cite{ness}.
However it appears as an increase (positive number) since 
$\vert d_j^{[i]}\vert$ is plotted on Fig. \ref{fig3}(b). 
Furthermore, the corresponding dimerisation is not simply the
superposition of the polaron and the
soliton---the two defects interact strongly together.
For example, the dimerisation pattern for an electron at position 
$i\approx25$ is characteristic of a merging of both the polaron and 
soliton.
Such mechanisms, leading to strong lattice distortions, would be 
expected to affect the transmission through the molecular 
wires---effects that we now consider.

\begin{figure}[p]
\psfig{figure=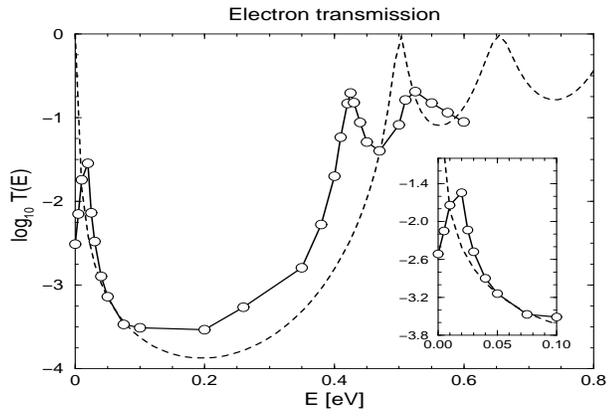,width=8cm,height=5.5cm}
\narrowtext
\caption{
\label{fig5}
Transmission probability $T(E)$ versus the energy $E$ for a molecular 
wire of length $N=99$.
{\it Dashed line}: transmission through a rigid chain containing a soliton 
defect (see the sharp peak at $E=0$ corresponding to the soliton state 
resonance).
{\it Solid line with circles}: transmission when the $e$-ph coupling is 
included.
The delocalisation of the soliton and the polaron-soliton 
interaction broaden the resonance peak at mid-gap into a ``mini-band''
(shown in the inset, enlarged low-energy region).
The polaron resonance peak appears around $E\approx 0.42{\rm \ eV}$.}
\end{figure}

We define an effective total transmission 
probability $T(E)$ arising from the contribution of the different 
outgoing channels as 
$T(E)\equiv\sum_a T_{ab}(\epsilon_{fin},\epsilon_{in})\ 
\delta(\epsilon_{fin}+\sum_\lambda n_\lambda\hbar\omega_\lambda-
\epsilon_{in})$
(see Fig. \ref{fig5}).
Without $e$-ph coupling, the transmission curve presents the usual
features: resonances at energies corresponding to the one-electron
levels of the molecule with almost perfect transmission, and
strong suppression of the transmission in the band-gap of the molecule
for which propagation occurs by tunneling.
The soliton resonance (at mid-gap) is much
narrower than the other resonances. 
This is to be expected since the corresponding one-electron (soliton) 
state is much more localized than the other states.
With $e$-ph interaction, we have already shown that a polaron can
be formed in the molecule and that the soliton position and width
are strongly modified.
Such mechanisms affect the transmission in the following way:
(i) the delocalization and deformation of the soliton broaden but lower
the resonance peak at mid-gap, 
(ii) for larger injection energies, the formation of the 
polaron effectively reduces the apparent band-gap, and a polaron resonance
peak appears (at $E\approx 0.42$ eV for the $N=99$ wire) inside the original 
gap. This effective band-gap reduction correlatively increases the 
transmission in the tunneling regime ({\it i.e.} for injection energies in 
the gap).

\section{Conclusion}

The results presented above demonstrate the complexity of electronic 
transport through one-dimensional atomic-scale wires.
We have illustrated the mechanisms of electron transport in molecular 
wires when $e$-ph interactions are included.
Owing to $e$-ph coupling, polarons can be formed inside the
molecule. Polaron propagation is the main mechanism of transport
through perfectly dimerised (semiconducting) molecules.
The presence of a mid-gap state associated to a soliton defect
in the middle of the molecule involves different mechanisms
for the transport.
For injection energies around the mid-gap state, the
delocalisation and deformation of the soliton is the main
mechanism for electron transfer.
With larger injection energies (still inside the energy band-gap 
of the molecule), a virtual polaron can be formed. 
The transport is then associated with more complicated mechanisms 
involving the interaction of both polaron and soliton.
However in most cases, the effective reduction of the
band-gap due to polaron formation, the delocalization and
deformation of the soliton and the polaron-soliton interaction
increase the transmission in the tunneling regime.
Although we studied a model derived from organic molecules, there is
good reason to believe that the $e$-ph coupling in other
one-dimensional nanoscale wires is likely to give rise to similar
phenomena.  
In the case of carriers injected into dangling-bond lines on the Si(001) 
surface \cite{dbwire}, polaron states which are in many respects 
analogous to those in molecular wires are also formed \cite{bowler01}.  
Specifically, we expect the present results to be valid for any 
atomic-scale wire in which there is a degeneracy between two different 
values of  some order parameter (the dimerisation in the molecular case, 
the surface dimer buckling in the dangling bond line) which are related 
by a discrete symmetry and strongly coupled to the electrons.

\acknowledgments

We acknowledge financial support from the U.K. EPSRC (Grant GR/M09193).

\end{multicols}

\end{document}